\documentclass[prd,aps,showpacs,nofootinbib,superscriptaddress,amsmath,amssymb,eqsecnum,preprint,tightenlines]{revtex4}

\usepackage{graphicx}
\usepackage{bm} 
\usepackage{epsfig}
\usepackage[latin1]{inputenc}
\usepackage{float,amsmath}

\def\lsim{\mathrel{\rlap{\lower4pt\hbox{$\sim$}}
    \raise1pt\hbox{$<$}}}                
\def\gsim{\mathrel{\rlap{\lower4pt\hbox{$\sim$}}
    \raise1pt\hbox{$>$}}}                
\newcommand{\beq}{\begin{equation}}
\newcommand{\eeq}{\end{equation}}
\newcommand{\bqa}{\begin{eqnarray}}
\newcommand{\eqa}{\end{eqnarray}}

\begin{document}


\title{Matching pre-equilibrium dynamics and viscous hydrodynamics}

\author{Mauricio Martinez}
\affiliation{
Helmholtz Research School and Otto Stern School\\
  Goethe-Universit\"at Frankfurt am Main\\
  Ruth-Moufang-Str.\,1, D-60438 Frankfurt am Main, Germany
}
\author{Michael Strickland}
\affiliation{
Physics Department, Gettysburg College, Gettysburg, \\
Pennsylvania 17325, USA\\
  \vspace{5mm}
}

\begin{abstract}
{
We demonstrate how to match pre-equilibrium dynamics of a 0+1 dimensional quark gluon plasma to 2nd-order viscous
hydrodynamical evolution.  The matching allows us to specify the initial values of the energy density and shear
tensor at the initial time of hydrodynamical evolution as a function of the lifetime of the pre-equilibrium period.  
We compare two models for the pre-equilibrium quark-gluon plasma, longitudinal free streaming and collisionally-broadened 
longitudinal expansion, and present analytic formulas which can be used to fix the necessary components of the
energy-momentum tensor.  The resulting dynamical models can be used to assess the effect of pre-equilibrium
dynamics on quark-gluon plasma observables.  Additionally, we investigate the dependence of entropy production on 
pre-equilibrium dynamics and discuss the limitations of the standard definitions of the non-equilibrium 
entropy.
}
\end{abstract}
\pacs{24.10.Nz, 25.75.-q, 12.38.Mh, 02.30.Jr} 
\maketitle

\section{Introduction}
\label{sec:intro}

The goal of relativistic heavy-ion experiments is to produce and characterize the thermodynamical and transport properties of matter 
at extremely high temperatures. In such collisions experimentalists expect to produce a new high-temperature state of matter that 
is a deconfined plasma of quarks and gluons (QGP).  Data from the Relativistic Heavy Ion Collider (RHIC) 
are consistent with the formation of a thermalized
QGP that exhibits strong transverse collective flow.  However, there are still many unresolved issues. One of them is to determine if 
the formed plasma is (nearly) isotropic and thermal at early times $\tau\sim 1-2$ fm/c.  Hydrodynamics
predictions for collective flow of the matter are consistent with RHIC data using thermalization times in
the range $\tau\sim 0.5-2$ fm/c \cite{Huovinen:2001cy,Hirano:2002ds,Tannenbaum:2006ch,Kolb:2003dz,Luzum:2008cw};
however, many outstanding questions remain.
One of the chief uncertainties in hydrodynamical \-mo\-de\-ling is the proper initial conditions to use when integrating the
resulting partial differential equations.  The initial conditions required are the energy density profile, ${\cal E}$, the components of the fluid four-velocity, $u^\mu$, and the stress tensor $\Pi^{\mu\nu}$ at an initial time $\tau_{\rm hydro}$. 

In this work we demonstrate how to determine the initial energy density ${\cal E}$ and shear 
$\Pi$ in a 0+1-dimensional model.  We introduce a pre-equilibrium period in which the system develops a
local momentum-space anisotropy owing to the longitudinal expansion of the matter.  After this period
we evolve the system using second-order viscous hydrodynamics with initial conditions consistent with the 
pre-equilibrium evolution of the matter.
To frame the discussion we introduce two proper time scales: (1) the parton formation time, $\tau_0$, which is the time after which 
coherence effects in the nuclear wave function for the hadrons can be ignored and partons can be thought of as liberated; and (2) the 
time at which modeling of the system using viscous hydrodynamics, $\tau_{\rm hydro}$ begins.
During the pre-equilibrium stage, $\tau_0 < \tau < \tau_{\rm hydro}$, longitudinal expansion of the matter along the beam axis makes 
the system colder along the longitudinal direction than in the transverse direction, $\langle p_L^2 \rangle <  \langle p_T^2 \rangle$
~\cite{Baier:2000sb} corresponding to a nonvanishing plasma shear $\Pi$.  

This paper extends previous work in which we introduced pre-equilibrium interpolating models
\cite{Mauricio:2007vz, Martinez:2008di,Martinez:2008mc} to 
calculate the dependence of high-energy dilepton and photon production on the plasma isotropization 
time~\cite{Mauricio:2007vz, Martinez:2008di, Martinez:2008mc, Schenke:2006yp, Bhattacharya:2008up, Bhattacharya:2008mv,Bhattacharya:2009sb,Ipp:2007ng, Ipp:2009ja}.  
In the previous analyses the
pre-equilibrium stage was matched at late times to isotropic ideal hydrodynamical expansion.  
Here we show how to determine the shear $\Pi$ and energy density $\cal E$ at a proper-time 
$\tau_{\rm hydro}$ given a model
for the evolution of the microscopic anisotropy of the plasma, $\xi = \frac{1}{2} 
\langle p_T^2 \rangle/\langle p_L^2\rangle-1$, where $p_{T}$ and $p_{L}$ are 
the transverse and longitudinal momenta of the particles in the plasma, respectively.
This is done by matching to the corresponding pressure anisotropy, 
$\Delta\equiv{\cal P}_T/{\cal P}_L - 1$, and energy density, ${\cal E}$.  Once this matching
is performed one can solve the second-order viscous hydrodynamical differential 
equations to determine the further time evolution of the system.  

\begin{figure*}[ht]
\includegraphics[width=16cm]{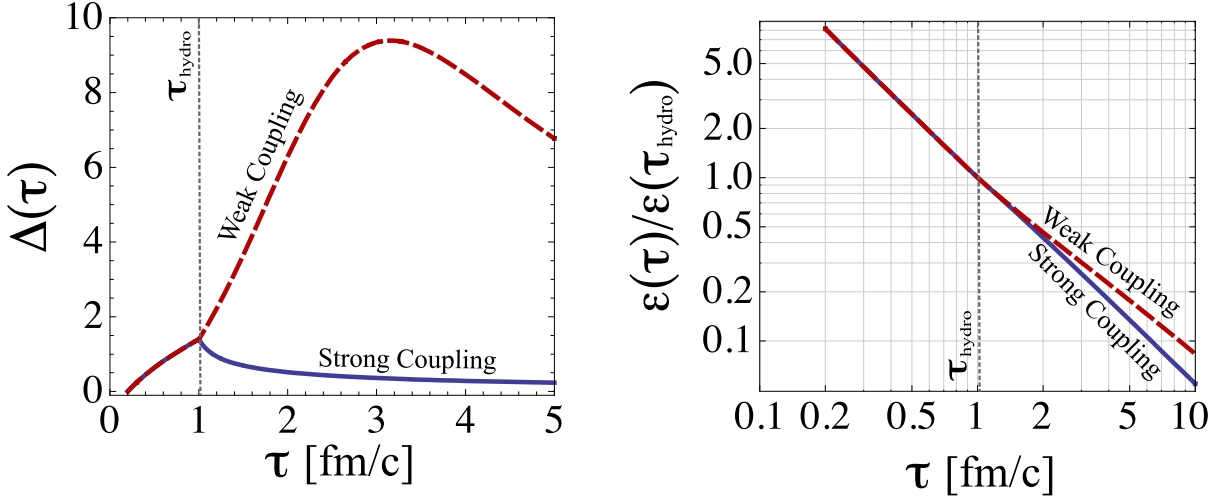}
\caption{Time evolution of the pressure anisotropy $\Delta$ (left) and energy density (right) 
for the case of collisionally broadened pre-equilibrium dynamics.  In this plot we have fixed $\tau_{\rm hydro}=1$
fm/c.  The dashed line (red) is the case of weakly coupled transport coefficients and the solid (blue) line is the case of
strongly coupled transport coefficients.}
\label{hydrocartoon}
\end{figure*}

In Fig.~\ref{hydrocartoon} we show the time
evolution of the pressure anisotropy and energy density assuming $\tau_{\rm hydro}=1$ fm/c
resulting from the models described herein.  As shown in Fig.~\ref{hydrocartoon} (left)
the magnitude of $\Delta$ is larger in the weakly coupled case starting from the same initial
pressure anisotropy at $\tau=1$ fm/c.  Figure \ref{hydrocartoon} (right) shows the typical
time evolution of the energy density using our matching.  As shown in this figure
in the weakly coupled case the 0+1-dimensional plasma lifetime is increased owing to the 
larger shear viscosity.  In the body of the text we show how such models are derived and
how, specifically, the matching at $\tau_{\rm hydro}$ is performed.  The resulting models
can be used as input to predict the effect of the pre-equilibrium period on QGP observables
such as dilepton and photon production, heavy quark screening, etc.

The work is organized as follows. In Sec.~\ref{vischydro}, we introduce the second-order
viscous hydrodynamics differential equations.  In Sec.\ref{pre-equilibrium}, we introduce the 
two models for the pre-equilibrium evolution and discuss how to determine the initial
conditions necessary to integrate the viscous hydrodynamics differential equations.
In Sec.~\ref{entropyprod}, making use of the resulting dynamical evolution, we calculate 
entropy production as a function of $\tau_{\rm hydro}$. 
In Sec.~\ref{conclusions}, we present our conclusions and outlook. 

\section{0+1 Viscous hydrodynamics}
\label{vischydro}

We consider a 0+1-dimensional system expanding in a boost-invariant manner along the longitudinal (beam-line)
direction with a uniform energy density in the transverse plane. In terms of proper time, $\tau =
\sqrt{t^2 -z^2}$, and space-time rapidity, $\zeta = {\rm arctanh}(z/t)$, the
second-order viscous hydrodynamic equations are given by \cite{Muronga:2003ta,Baier:2007ix}:
\begin{subequations}
\label{visceqs}
\begin{align}
\partial_\tau {\cal E}&=-\frac{{\cal E}+{\cal P}}{\tau}+\frac{\Pi}{\tau} \, ,
\label{0+1eqe}\\
\partial_\tau \Pi &= -\frac{\Pi}{\tau_\pi}
+\frac{4 \eta}{3\, \tau_\pi \tau}-\frac{4}{3\, \tau} \Pi
-\frac{\lambda_1}{2\,\tau_\pi\,\eta^2} \left(\Pi\right)^2 \, ,
\label{0+1eqp}
\end{align}
\end{subequations}
where ${\cal E}$ is the fluid energy density, ${\cal P}$ is the fluid pressure, 
$\Pi \equiv \Pi^\zeta_\zeta$ is the $\zeta\zeta$ component 
of the fluid shear tensor, $\eta$ is the fluid shear viscosity, $\tau_\pi$ is the shear 
relaxation time, and $\lambda_1$ is a coefficient that arises in complete second-order 
viscous hydrodynamical equations in either the strongly \cite{Baier:2007ix,Bhattacharyya:2008jc} or the weakly coupled limit \cite{Muronga:2003ta,York:2008rr,Betz:2008me}.

To solve these coupled differential equations it is necessary to specify the initial conditions and the
equation of state that relates the energy density and the pressure through ${\cal P} = {\cal P}({\cal E})$. 
For 0+1-dimensional viscous hydrodynamics, one must specify the energy density and $\Pi$ at the initial time, that is, 
${\cal E}_{\rm hydro} \equiv {\cal E}(\tau_{\rm hydro})$ and $\Pi_{\rm hydro} \equiv \Pi(\tau_{\rm hydro})$, where $\tau_{\rm hydro}$ 
is the proper time at which one begins to solve the differential equations.

In the following analysis we assume an ideal equation of state, in which case we have
\beq 
\label{eqstate}
{\cal P} = \frac{N_{\rm dof}\, \pi^2}{90} T^4\, ,
\eeq
where for quantum chromodynamics with $N_c$ colors and $N_f$ quark flavors,
$N_{\rm dof} = 2 (N_c^2-1) + 7 N_c N_f/2$, which, for $N_c=3$ and $N_f=2$ 
is $N_{\rm dof} = 37$.  The general method used here, however, can easily be extended
to a more realistic equation of state. 

In the conformal limit the trace of the four-dimensional stress tensor vanishes requiring
${\cal E} = 3 {\cal P}$ which, using Eq.~(\ref{eqstate}), allows us to write compactly
\beq
{\cal E} = (T/\gamma)^4, \hspace{0.5cm} \text{with}\hspace{0.5cm}\gamma \equiv \left( \frac{30}{\pi^2 N_{\rm dof}} 
\right)^{1/4} \,.
\label{energyideal}
\eeq
Likewise we can simplify the expression for the equilibrium entropy density, ${\cal S}$, using the thermodynamic relation 
$T {\cal S} = {\cal E} + {\cal P}$ to obtain ${\cal S} = 4 {\cal E} / 3 T$ or, equivalently,
\beq
{\cal S} = \frac{4}{3 \gamma}\, {\cal E}^{3/4} \, .
\label{entropyideal}
\eeq
Note that for a system out of equilibrium the full nonequilibrium entropy is modified compared to (\ref{entropyideal}). Using kinetic theory it is possible to show that the entropy current receives corrections at second order in gradients~\cite{GLW}. In the original approach of Israel and Stewart~\cite{Israel:1976tn,Israel:1979wp}, the non-equilibrium entropy is expanded
in a series in deviations from equilibrium and higher-order corrections are neglected. The 
Israel-Stewart (IS) ansatz for the nonequilibrium entropy is
\beq
\label{viscentropy}
{\cal S_{\rm noneq}^{\rm IS}}= {\cal S}-\frac{\beta_2}{2 T}\Pi_{\mu\nu}\Pi^{\mu\nu} \, ,
\eeq
where $\beta_2$ is an a priori unknown function that determines the importance of second-order modifications to
the entropy current.  The IS ansatz satisfies the second law of thermodynamics $\partial^\mu {\cal S}_\mu^{\rm IS}\geq 0$
and for massless particles described by a Boltzmann distribution function one finds $\beta_2=\tau_\pi/(2\eta)$.  
Recent analyses have shown that, including all relevant structures in the gradient expansion, the non-equilibrium entropy  
contains additional terms not present in the simple IS definition of the non-equilibrium entropy~\cite{Romatschke:2009kr,Loganayagam:2008is, Bhattacharyya:2008xc,Lublinsky:2009kv,Lublinsky:2007mm,El:2009vj}. 

\begin{table}[t]
\begin{tabular}{ | c || c | c | c |}
\hline
{\bf $\;$ Transport coefficient $\;$} & 
{\bf $\;$ Weakly-coupled QCD $\;$ } & 
{\bf $\;$ Strongly-coupled ${\cal N}=4$ SYM $\;$} \\ \hline
$\bar{\eta}\equiv \eta/{\cal S}$ & $\thicksim1/(g^4 \log g^{-1})$ & $1/(4 \pi)$\\ \hline
$\tau_\pi$ & $6 \bar{\eta}/T$ & $ 2 \bigl(2-\log 2\bigr) \bar\eta /T$ \\ \hline
$\lambda_1$ & $(4.1 \rightarrow 5.2)\,\bar{\eta}^2 {\cal S}/T$ &  2 $\bar{\eta}^2 {\cal S}/T$\\ \hline
\end{tabular}
\caption{Typical values of the transport coefficients for a weakly-coupled QGP 
\cite{York:2008rr,Arnold:2000dr,Arnold:2003zc} and 
a strongly coupled ${\cal N}=4$ SYM plasma \cite{Baier:2007ix,Bhattacharyya:2008jc}.}
\label{transcoeff}
\end{table}

When solving Eqs.~(\ref{0+1eqe}) and (\ref{0+1eqp}) it is important to recognize that 
the transport \-coe\-ffi\-ci\-ents depend on the temperature of the plasma and hence on the proper time. We summarize in 
Table~\ref{transcoeff} the values of the transport coefficients in the 
strong~\cite{Baier:2007ix,Bhattacharyya:2008jc} and weak~\cite{York:2008rr,Arnold:2000dr,Arnold:2003zc} coupling limits. 
We point out that in both cases the transport coefficients do not \-sa\-tis\-fy universal relations and therefore, their values can 
only be taken as estimates. In the weakly coupled case the QCD transport coefficients depend on the renormalization scale. In addition, 
higher-order corrections to some transport coefficients from finite-temperature perturbation theory show poor 
convergence~\cite{CaronHuot:2007gq,CaronHuot:2008uh}. At strong coupling, it has been shown recently that there are corrections for 
finite t'Hooft coupling \cite{Brigante:2007nu,Brigante:2008gz,Kats:2007mq,Natsuume:2007ty,Buchel:2008vz}. We take the preceding 
estimates in both coupling limits to get a qualitative understanding of what to expect in each regime.

In Table~\ref{transcoeff}, the reader should note that in the strong and weak coupling limit the coefficients $\tau_\pi$ and $\lambda_1$ are proportional to $\tau_\pi\propto \bar\eta / T$ and 
$\lambda_1\propto \bar\eta^2 {\cal S} / T$, respectively. This fact suggests that we can parametrize both coefficients as
\begin{subequations}
 \label{parametrization}
\begin{align}
 \tau_\pi & =  \frac{c_\pi\,\bar\eta}{T} \, ,\\
 \lambda_1 & =  c_{\lambda_1} \bar{\eta}^2 \biggl(\frac{{\cal S}}{T}\biggr)\, ,
\end{align}
\end{subequations}
where we have introduced the scaled shear viscosity 
\beq
\bar{\eta}\equiv \eta / {\cal S} \, .
\eeq 
In our analysis we assume that $\bar\eta$ is independent of time.
 
The dimensionless numbers $\bar\eta$, $c_\pi$ and $c_{\lambda_1}$ carry all of the information about the particular coupling limit we are considering. 
Using the ideal gas equation of state [Eqs. (\ref{energyideal}) and (\ref{entropyideal})], the parametrization~(\ref{parametrization}) of $\tau_\pi$ and $\lambda_1$ can be rewritten in terms of the energy density ${\cal E}$

\begin{subequations}
\label{parametrization2}
 \begin{align}
 \tau_\pi & =  \frac{c_\pi\,\bar\eta}{\gamma\, {\cal E}^{1/4}} \, ,\\  
 \lambda_1 & =  \frac{4}{3\gamma^2}\, c_{\lambda_1}\, \bar{\eta}^2 \, {\cal E}^{1/2} \, .
 \end{align}
\end{subequations}

We use the following values for the transport coefficients in the case of a weakly coupled QGP
\begin{equation}
\label{weaklimitvalues}
\bar\eta = \frac{10}{4 \pi}\, ,\hspace{0.5cm}c_\pi = 6\, ,\hspace{0.5cm}c_{\lambda_1} = \frac{9}{2} \, .
\end{equation}
For the strong coupling analysis, we use  
\begin{equation}
\label{stronglimitvalues}
\bar\eta = \frac{1}{4 \pi} \, , \hspace{0.5cm}c_\pi = 2\;(2 - \log 2) \, ,\hspace{0.5cm}c_{\lambda_1} = 2 \, .
\end{equation}


\subsection{Pressure anisotropy}
\label{sec:momentumanisotropybounds}
We introduce the dimensionless parameter $\Delta$, which measures the pressure anisotropy of the fluid as follows
\beq
\label{deltavisc}
\Delta \equiv \frac{{\cal P}_T}{{\cal P}_L} - 1 \, ,
\eeq
where ${\cal P}_T = (T^{xx} + T^{yy})/2$ and ${\cal P}_L = T^{zz} = - T^\zeta_\zeta$ are the effective transverse and longitudinal 
pressures, respectively.  If $\Delta=0$, the system is locally isotropic.  If $-1 < \Delta < 0$, the system has a local prolate 
anisotropy in momentum space and if $\Delta >0$, the system has a local oblate anisotropy in momentum space.

In the 0+1-dimensional model of viscous hydrodynamics one can express the effective transverse pressure as 
${\cal P}_T = {\cal P} + \Pi/2$ and the effective longitudinal pressure as ${\cal P}_L = {\cal P} - \Pi$.  Using these definitions for ${\cal P}_T$ and ${\cal P}_L$ and the ideal equation of state, we can rewrite Eq. (\ref{deltavisc}) as
\beq
\label{deltapi}
\Delta (\tau) = \frac{9}{2} \left(\frac{\Pi (\tau)}{{\cal E} (\tau) - 3\,\Pi (\tau)}\right) \, .
\eeq
In different limits we have
\begin{subequations}
 \begin{align}
  \label{deltapismall}
  \lim_{\Pi\ll {\cal E}}\,\Delta (\tau) &\approx \frac{9}{2}\frac{\Pi (\tau)}{{\cal E} (\tau)}\, , \\
  \label{deltapinegat}
  \lim_{\Pi \rightarrow -2{\cal E}/3}\,\Delta (\tau) &\approx -1\, ,\\
  \label{deltapiinf}
   \lim_{\Pi \rightarrow {\cal E}/3}\,\Delta (\tau) &\approx \infty .
 \end{align}
\end{subequations}
At the initial time $\tau=\tau_{\rm hydro}$, $\Delta_{\rm hydro} \equiv \Delta(\tau=\tau_{\rm hydro})$ is given by
\beq
\label{deltahydro}
\Delta_{\rm hydro} = \frac{9}{2} \left(\frac{\Pi_{\rm hydro}}{{\cal E}_{\rm hydro} - 3\,\Pi_{\rm hydro}}\right) \, ,
\eeq
or, 

\beq
\label{pihydro}
 \Pi_{\rm hydro} = \frac{2\,\Delta_{\rm hydro}}{3\,+\,2\,\Delta_{\rm hydro}}\,{\cal E}_{\rm hydro}\, ,
\eeq

This expression allows one to find the initial condition for the shear tensor component $\Pi_{\rm hydro}$ knowing 
$\Delta_{\rm hydro}$ and ${\cal E}_{\rm hydro}$. This relation is the bridge that allows us to match the initial condition for the 
shear tensor $\Pi_{\rm hydro}$ from a pre-equilibrium period of the QGP.  The precise matching is described in 
Sect.~\ref{pre-equilibrium}, where we derive the connection between $\Delta$ defined in Eq.~(\ref{deltavisc}) and the $\xi$ parameter 
introduced in Ref.~\cite{Romatschke:2003ms}.


\section{0+1-dimensional model for a pre-equilibrium QGP}
\label{pre-equilibrium}

In this section we present two models for 0+1-dimensional nonequilibrium time evolution of the QGP: 0+1-dimensional free-streaming and 
0+1-dimensional collisionally broadening expansion. In each case below we will be required to specify a proper time dependence of the hard-momentum scale, $p_{\rm hard}$, and the microscopic anisotropy parameter, $\xi$, introduced in Ref.~\cite{Romatschke:2003ms}. Before proceeding, however, it is useful to note some general relations.  

We assume that any anisotropic distribution function can be obtained by taking an arbitrary isotropic distribution function 
$f_{\rm iso}(p)$ and stretching or squeezing it along one direction in momentum space to obtain an anisotropic distribution. This can 
be achieved with the following parametrization
\beq
f_{\rm aniso}({\bf p},\xi,p_{\rm hard})=f_{\rm iso}(\sqrt{{\bf p^2}+\xi({\bf p\cdot \hat{n}}){\bf^2}},p_{\rm hard}) \; ,
\label{eq:distansatz}
\eeq
where $p_{\rm hard}$ is the hard momentum scale, $\hat{\bf n}$ is the direction of the anisotropy,\footnote{Hereafter, we use 
$\hat{\bf n}=\hat{\bf e}_z$, where $\hat{\bf e}_z$ is a unit vector along the beam-line direction.} and $-1 < \xi < \infty$ is a parameter that reflects the strength and type of anisotropy. In general, $p_{\rm hard}$ is related to the average momentum in the partonic distribution function. The microscopic plasma anisotropy parameter $\xi$ is related to the average longitudinal and transverse momentum of the plasma partons via the relation \cite{Mauricio:2007vz,Martinez:2008di,Martinez:2008mc,Dumitru:2007hy}
\beq
\xi=\frac{\langle p_T^2\rangle}{2\langle p_L^2\rangle} - 1 \; .
\label{anisoparam}
\eeq
From this expression, one can see that for an oblate plasma $\langle p_T^2\rangle > 2\langle p_L^2\rangle$, then $\xi>0$.
In an isotropic plasma one has $\xi=0$, and in this case, $p_{\rm hard}$ can be 
identified with the plasma temperature $T$. 

We now show how to derive a general formula for the time evolution of the microscopic plasma anisotropy
$\xi$ that allows for a nonvanishing anisotropy of the plasma at the formation time followed by subsequent
dynamical evolution.  This is a straightforward
extension of the treatment presented in Ref.~\cite{Martinez:2008di} where it was assumed that the plasma was
isotropic at the formation time.

In most phenomenological approaches to QGP dynamics it is assumed that the distribution function at $\tau\sim\tau_0$ is isotropic, 
that is, $\xi (\tau=\tau_0)$=0. There is no clear justification for this assumption. In fact, 
in the simplest form of the Color Glass Condensate (CGC) model \cite{McLerran:1993ka}  the 
longitudinal momentum would initially be zero. 
This configuration corresponds to an extreme anisotropy with $\xi$ (or $\Delta$) being infinite in the initial state.
In the CGC framework to generate a nonzero longitudinal pressure it is necessary to include
the next-to-leading-order corrections to gluon production taking into account the effect of rapidity fluctuations 
and full three-dimensional gauge field dynamics. There has been progress toward the solution of this problem; however, it is still an 
open question (for recent advances in this area see Refs.~\cite{Lappi:2009mp,Rebhan:2008uj} and references therein).
We also note that, 
taking into account the finite longitudinal width of the nuclei, studies have shown that it may even be possible for the
initial plasma anisotropy to be prolate at the formation time \cite{Jas:2007rw}.

Here we assume, quite generally, that the microscopic anisotropy parameter, $\xi$, at the formation time
takes on an arbitrary value between -1 and $\infty$ given by $\xi_0$.  The initial anisotropy $\xi_0$ can be evaluated using 
Eq.~(\ref{anisoparam}) giving
\bqa
\label{initialxi0}
\xi_0=\frac{1}{2}\frac{\langle p_T^2\rangle_0}{\;\langle p_L^2\rangle_0}-1 \, .
\eqa
Writing the longitudinal momentum as
\beq
\label{ansatzpl}
\langle p_L^2\rangle\;=\;\langle p_L^2\rangle_0\;+\;\langle \delta p_L^2\rangle \, ,
\eeq
and using the fact that, in the case of 0+1 dynamics, the average transverse momentum is constant
\beq
\langle p_T^2\rangle\;=\;\langle p_T^2\rangle_0 \, ,
\eeq
we can rewrite the general expression for $\xi$ given in Eq.~(\ref{anisoparam}) as
\beq
\label{initialxi1}
\xi = \frac{\xi_0+1}{1+\frac{\langle \delta p_L^2\rangle}{\langle p_L^2\rangle_0}}-1.
\eeq
Finally, we can parametrize the time dependence of the plasma's average longitudinal momentum squared as
\begin{equation*}
\langle p_L^2\rangle\sim \langle p_L^2\rangle_0 \left(
\frac{\tau_0}{\tau}\right)^{\delta}.
\end{equation*}
Comparing with Eq.~(\ref{ansatzpl}), we obtain
\beq
\frac{\langle \delta p_L^2\rangle}{\langle p_L^2\rangle_0}= \biggl(\frac{\tau_0}{\tau}\biggr)^{\delta}-1
\eeq
Inserting this into Eq.~(\ref{initialxi1}), we  have
\beq
\label{generalxi}
\xi(\xi_0,\tau,\tau_0)= (\xi_0+1)\biggl(\frac{\tau}{\tau_0}\biggr)^{\delta}-1
\eeq
This expression holds for both of the  0+1-dimensional pre-equilibrium scenarios studied in this work.  In
the case of longitudinal free streaming  we have $\delta = 2$, and in the case of collisional broadening  we have 
$\delta = 2/3.$\footnote{Owing to the assumption of no dynamics in the transverse plane, collisional broadening can
only increase the longitudinal momentum in 0+1 dimensions.} 
There are other possibilities for the values of this exponent associated with the bending caused by growth of the chromoelectric and 
chromomagnetic fields at early times of the collision.  We refer the reader to Sec. III of Ref. \cite{Martinez:2008di} for an extended discussion of the time dependence
of the scaling coefficient $\delta$.

In a comoving frame, the energy density and pressure components can be determined by evaluating the components of the stress-energy 
tensor,
\beq
\label{stresstensoraniso}
T^{\mu\nu} = \int \frac{d^3 {\bf p}}{(2 \pi)^3} \frac{p^\mu p^\nu}{p^0 } f({\bf p},p_{\rm hard}) \, . 
\eeq
Using the ansatz, Eq.~(\ref{eq:distansatz}), for the anisotropic distribution function and making an appropiate change of variables, 
one can show that the local energy density ${\cal E}$ and the transverse and longitudinal pressures ${\cal P}_T$ and ${\cal P}_L$ are
\begin{subequations}
\label{momentsanisotropic}
\begin{align}
\label{energyaniso}
{\cal E}(p_{\rm hard},\xi) &= T^{00} \;= \frac{1}{2}\Biggl(\frac{1}{1+\xi}
+\frac{\arctan\sqrt{\xi}}{\sqrt{\xi}} \Biggr) {\cal E}_{\rm iso}(p_{\rm hard}) \; , \\ \nonumber
&={\cal R}(\xi)\,{\cal E}_{\rm iso}(p_{\rm hard})\, ,\\
\label{transpressaniso}
{\cal P}_T(p_{\rm hard},\xi) &= \frac{1}{2}\left( T^{xx} + T^{yy}\right)\, , \\ \nonumber
&= \frac{3}{2 \xi} 
\left( \frac{1+(\xi^2-1){\cal R}(\xi)}{\xi + 1}\right)
 {\cal P}_T^{\rm iso}(p_{\rm hard}) \, , \\ \; 
\label{longpressaniso}
{\cal P}_L(p_{\rm hard},\xi) &= T^{zz}\, ,  \\ \nonumber
&= \frac{3}{\xi} 
\left( \frac{(\xi+1){\cal R}(\xi)-1}{\xi+1}\right) {\cal P}_L^{\rm iso}(p_{\rm hard}) \; ,
\end{align}
\end{subequations}
where ${\cal P}_T^{\rm iso}(p_{\rm hard})$ and ${\cal P}_L^{\rm iso}(p_{\rm hard})$ are the isotropic transverse and longitudinal 
pressures and ${\cal E}_{\rm iso}(p_{\rm hard})$ is the isotropic energy density.~\footnote{We point out that, in general, one cannot 
identify ${\cal P}_T^{\rm iso}(p_{\rm hard})$, ${\cal P}_L^{\rm iso}(p_{\rm hard})$ and ${\cal E}_{\rm iso}(p_{\rm hard})$ with their 
equilibrium counterparts, unless one implements the Landau matching conditions. In Appendix~\ref{Ap:A}, we show an explicit example 
where the Landau matching conditions are implemented.} 
The function ${\cal R}(\xi)$ is given by
\bqa
{\cal R}(\xi)\;=\;\frac{1}{2}\biggl[\;\frac{1}{1+\xi}+\frac{\text{arctan}\sqrt{\xi}}{\sqrt{\xi}}\;
\biggr] ,
\eqa 
and in Eq.~(\ref{momentsanisotropic}) it is understood that $p_{\rm hard}=p_{\rm hard}(\xi_0,\tau,\tau_0,\delta)$ and 
$\xi=\xi(\xi_0,\tau,\tau_0,\delta)$.

Note that for a conformal system the tracelessness of the stress-energy tensor $T^\mu_\mu$=0 implies 
${\cal E} = 2{\cal P}_T + {\cal P}_L$. This condition is satisfied by Eqs.~(\ref{momentsanisotropic}) for any anisotropic 
distribution function (Eq.~\ref{eq:distansatz}) because for an isotropic conformal state 
${\cal P}_{T,L}^{\rm iso}={\cal E}_{\rm iso}/3$.


\subsection{0+1-dimensional free streaming expansion}
\label{freestr}

In the free streaming (f.s.) case, the distribution function is a solution of 
the collisionless Boltzmann equation
\bqa
p\,\cdot\,\partial_x\,f_{\rm f.s.}(p,x)=0 \;,
\label{eq:freestreamb}
\eqa
Consider for simplicity that the one-particle distribution function is isotropic at the formation time, $\tau=\tau_0$.
\begin{equation}
f_{\rm f.s.}(p,x)\Biggr|_{\tau=\tau_0}=f\Biggl(\frac{\sqrt{p_T^2+p_L^2}}{p_{\rm hard}}\Biggr) \; ,
\end{equation}
where $p_T$ is the transverse momentum, $p_L$ is the longitudinal momentum, and $p_{\rm hard}$ is the hard momentum scale at $\tau_0$. 
The hard momentum scale for particles undergoing 0+1-dimensional free streaming expansion is constant in time. 
In comoving coordinates the general solution for free streaming expansion in 0+1-dimensional expansion can be written as
\bqa
\label{fsfunction}
f_{\rm f.s.}(p,x) = 
  f\Biggr(\frac{p_T}{p_{\rm hard}}\sqrt{1+\frac{\tau^2}{\tau_0^2}\sinh^2\,(y-\zeta)}\,\Biggr) \; ,
\label{eq:fssol2}
\eqa
where $y$ is the momentum-space rapidity, $\tau$ is the proper time, and 
$\zeta$ is the space-time rapidity.

Using the free streaming distribution function, Eq.~(\ref{eq:fssol2}), it is possible to show that the average longitudinal and 
transverse momentum-squared values are given by \cite{Martinez:2008di, Baier:2000sb}
\begin{subequations}
 \begin{align}
  \frac{1}{2} \langle p_T^2 \rangle_{\rm f.s.} &\propto p_{\rm hard}^2 \; ,\\
  \langle p_L^2 \rangle_{\rm f.s.} &\propto p_{\rm hard}^2 \, \frac{\tau_0^2}{\tau^2} \; .
 \end{align}
\end{subequations}
Inserting these expressions into the general expression for $\xi$ given in Eq.~(\ref{anisoparam})
one obtains $\xi_{f.s.}(\tau) = (1+\xi_0)(\tau/\tau_0)^2-1$. Therefore, $\delta = 2$ in Eq.~(\ref{generalxi}).

One can also determine the temporal evolution of the energy density for the 0+1-dimensional free streaming case
using Eq.~(\ref{momentsanisotropic}).  The resulting temporal evolution of the relevant variables is~\cite{Martinez:2008di}
\begin{subequations}
\begin{align}
\xi_{\rm f.s.}(\tau) &= (1+\xi_0)\bigl(\tau/\tau_0\bigr)^2 - 1 \; , \label{fsXIeq}\\
p_{\rm hard}(\tau) &= p_{\rm hard}  \; , \label{fsTeq} \\
{\cal E}(\tau) &={\cal R}(\xi_{\rm f.s.})\; \biggl(\frac{p_{\rm hard}}{\gamma}\biggr)^4 . \label{fsEeq} 
\end{align}
\label{fslimit}
\end{subequations}


\subsection{0+1-dimensional collisionally broadening expansion}
\label{collbroad}

In the bottom up scenario~\cite{Baier:2000sb}, it was shown that, even at early times after the nuclear impact, 
elastic collisions between the liberated partons will cause a broadening of the longitudinal momentum of the particles compared to 
the noninteracting, free-streaming case. During the first stage of the bottom-up scenario, when $1\ll Q_s\tau\ll\alpha_s^{3/2}$, 
the initial hard gluons have typical momentum of order $Q_s$ and occupation number of order $1/\alpha_s$. Owing to the fact that the 
system is initially expanding at the speed of light in the longitudinal direction the gluon number density drops like 
$n_g \sim Q_s^3/(\alpha_s Q_s\tau)$. If there were no interactions, this expansion would be equivalent to 0+1-dimensional free 
streaming and the longitudinal momentum $p_L$ would scale like $1/\tau$. However, when elastic $2\leftrightarrow 2$ collisions of hard 
gluons are taken into account \cite{Baier:2000sb}, the ratio between the longitudinal momentum $p_L$ and the typical transverse 
momentum of a hard particle $p_T$ decreases as
\begin{equation}
\label{ptbroadbottom}
\frac{\langle p_L^2 \rangle}{\langle p_T^2 \rangle} \propto (Q_s\tau)^{-2/3} \; .
\end{equation}
This implies that for a collisionally-broadened (c.b.) plasma, $\xi_{\rm c.b.}=(1+\xi_0)(\tau / \tau_0)^{2/3}-1$, 
implying $\delta = 2/3$ in Eq.~(\ref{generalxi}).

The temporal evolution of $\xi$, $p_{\rm hard}$ and ${\cal E}$ for the case of 
0+1 collisionally-broadened expansion is~\cite{Martinez:2008di}
\begin{subequations}
\begin{align}
\xi_{\rm c.b.}(\tau) &= (1+\xi_0)\bigl(\tau/\tau_0\bigr)^{2/3} - 1 \; , \label{cbXIeq}\\
p_{\rm hard}(\tau) &= (p_{\rm hard})_0\;\bigl(\tau_0/\tau\bigr)^{2/9} \; , \label{cbTeq} \\
{\cal E}(\tau) &={\cal R}(\xi_{\rm c.b.})\;\biggl(\frac{p_{\rm hard}}{\gamma}\biggr)^4 . 
\label{cbEeq} 
\end{align}
\label{cblimit}
\end{subequations}
%


\subsection{Relation between $\Delta$ and $\xi$}
\label{app:xideltarelation}

In this section, we derive the relation between the pressure anisotropy parameter, $\Delta$, introduced in 
Sec.~\ref{sec:momentumanisotropybounds}, and the microscopic anisotropy parameter, $\xi$. 
Combining Eqs. (\ref{transpressaniso}) and (\ref{longpressaniso}) and using 
$\mathcal P_T^{\rm iso} = \mathcal P_L^{\rm iso} = {\cal E}_{\rm iso}/3$, we obtain the following expression for 
$\Delta$ 
\begin{eqnarray}\nonumber
\Delta (\xi) &=& \frac{\mathcal P_T(\xi)}{\mathcal P_L(\xi)}-1\, ,\\ 
&=& \frac{1}{2} \left(\xi -3\right) + 
\xi \left(  (1+\xi)\frac{\text{atan}\sqrt{\xi}}{\sqrt{\xi}} - 1 \right)^{-1}\, .
\label{xideltamatch}
\end{eqnarray}
The evolution of $\Delta$ during the pre-equilibrium stage will depend on the kind of model for that stage, that is, either free 
streaming or collisionally-broadened expansion. For small and large values of $\xi$ 
\begin{subequations}
\begin{align}
\label{smallxilimit}
 \lim_{\xi \rightarrow 0} \Delta &= \frac{4}{5} \xi + {\cal O}(\xi^2) \, ,\\
\label{largexilimit}
\lim_{\xi \rightarrow \infty} \Delta &= \frac{1}{2} \xi + {\cal O}(\sqrt{\xi}) \, .
\end{align}
\end{subequations}
If one uses Eq.~(\ref{deltapismall}) together with Eq.~(\ref{smallxilimit}), $\xi$ can be related with the shear viscous tensor during the viscous regime as~\footnote{An alternative derivation of this relation from kinetic theory is presented in Appendix~\ref{Ap:C}.}
\beq
\label{smallxiviscous}
\xi= \frac{45}{8}\frac{\Pi}{{\cal E}}+{\cal O}(\Pi^2)\,.
\eeq
Note that if one uses the Navier-Stokes value of the shear tensor $\Pi_{\rm NS}=4\eta/(3\tau)$ in the last relation, the anisotropy parameter can be expressed as~\cite{Asakawa:2006jn}
\beq
\xi_{\rm NS} = \frac{10}{T \tau}\frac{\eta}{{\cal S}}+{\cal O}(\Pi^2_{\rm NS})\,.\eeq
%
 

\subsection{Matching the initial conditions}
\label{matching}
We now match the general evolution of $\xi$ from Eq.~(\ref{generalxi}) at an intermediate $\tau_{\rm hydro}$ and use this to fix the 
initial shear tensor that should be used in the viscous evolution.
From Eq. (\ref{generalxi}), the anisotropy parameter takes a nonvanishing value at $\tau = \tau_{\rm hydro}$, 
\beq
\xi_{\rm hydro}\equiv\xi (\tau=\tau_{\rm hydro})=(1+\xi_0)\biggl(\frac{\tau_{\rm hydro}}{\tau_0}\biggr)^\delta-1\, .
\eeq
Once $\xi_{\rm hydro}$ is known, the initial pressure anisotropy $\Delta_{\rm hydro}$ and initial energy density 
${\cal E}_{\rm hydro}$ can be determined using Eqs.~(\ref{xideltamatch}) and~(\ref{energyaniso}) respectively.  It is then 
straightforward to determine $\Pi_{\rm hydro}\equiv\Pi (\tau=\tau_{\rm hydro})$ 
owing to relation~(\ref{pihydro}), that is, 
\beq
\Pi_{\rm hydro} \,\equiv\,  \Pi\,(\tau=\tau_{\rm hydro}) = \frac{2}{3}\,\biggl(\frac{\Delta_{\rm hydro}}{3\,+\,2\,\Delta_{\rm hydro}}\biggr)\,{\cal E}_{\rm hydro}\,.
\label{pideltamatch}
\eeq 
This expression together with ${\cal E}_{\rm hydro}$ gives the full set of initial conditions necessary to solve the 0+1-dimensional 
viscous hydrodynamics equations~(\ref{visceqs}). Note that, by construction, the initial conditions do not depend on the particular 
coupling regime we are interested in. This is because at leading order the coupling constant cancels out in the case of a collisionally-broadened expansion and in the case of free-streaming it is assumed that there is only free expansion.  As a result
$\Pi_{\rm hydro}$ and ${\cal E}_{\rm hydro}$ depend only on the type of pre-equilibrium scenario considered through the exponent $\delta$. 

\section{Temporal evolution including pre-equilibrium dynamics}
\label{results}

\begin{figure*}[t]
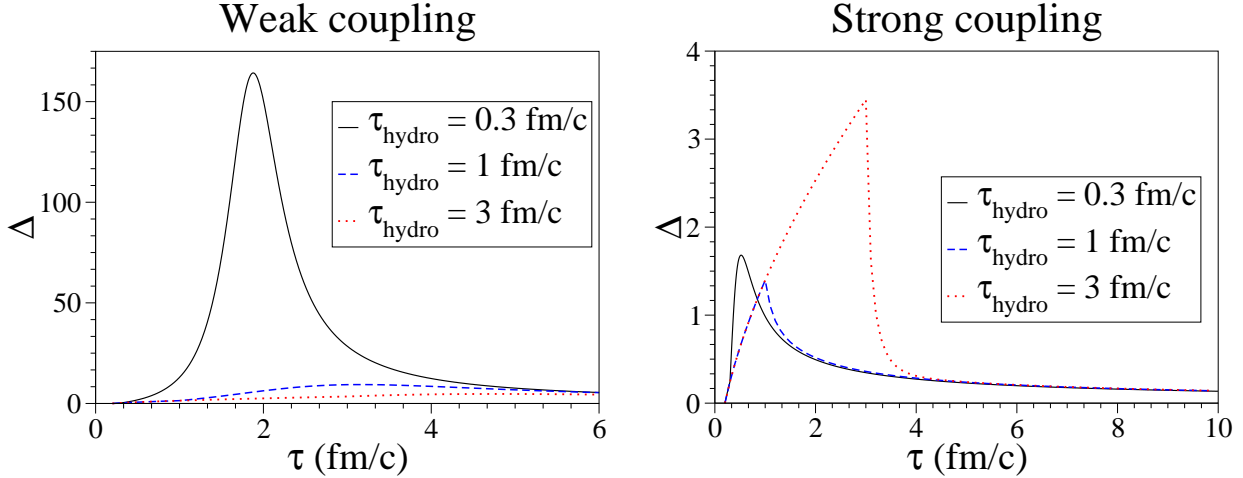

\includegraphics[scale=0.31]{2a.eps}\hspace{5mm}
\includegraphics[scale=0.31]{2b.eps}
\caption{Temporal evolution of the pressure anisotropy parameter $\Delta$ for three values of $\tau_{\rm hydro} \in 
\{0.3,1,3\}$ fm/c. We use $\tau_0$ = 0.3 fm/c, the initial temperature at the parton formation time $p_{\rm hard} = T_0$ = 0.35 GeV,
and an initial value for $\xi_0$=0. Both plots assume the collisionally-broadened scenario, and the transport coefficients during the 
viscous period correspond to the weak coupling (right) and strong (left) coupling regimes.}
\label{tempevol}
\end{figure*}

In Fig.~\ref{tempevol}, we show the complete temporal evolution of the pressure anisotropy $\Delta(\tau)$, starting from a 
pre-equilibrium period and matching at $\tau_{\rm hydro}$ to viscous hydrodynamical evolution. In the plot we show three assumed 
values of $\tau_{\rm hydro}$. The initial conditions for the strong and weak coupling cases are assumed to be the same in both panels. 
During the pre-equilibrium case $\tau_0\leq\tau<\tau_{\rm hydro}$, $\Delta(\tau)$ is determined via its relation to $\xi(\tau)$ 
specified in Eq.~(\ref{xideltamatch}).  In Fig.~\ref{tempevol} we have shown the case where $\xi$ evolves in the collisionally 
broadened scenario, that is, $\delta=2/3$. The matching from pre-equilibrium dynamics to viscous evolution occurs at 
$\tau_{\rm hydro}$, where, owing to the longitudinal expansion of the plasma, a nonvanishing value of $\xi$ is generated. Using 
Eqs.~(\ref{energyaniso}), (\ref{xideltamatch}), and (\ref{pideltamatch}) we use the value of $\xi(\tau_{\rm hydro})$ to determine 
the initial values of the energy density and shear necessary for integration of the viscous hydrodynamical differential equations. 
From $\tau_{\rm hydro}\leq\tau\leq\tau_{\rm fo}$, $\Delta$ is determined using Eq.~(\ref{deltapi}). It should be understood that 
during this period of the evolution the energy density and shear are the solutions of the viscous hydrodynamical differential 
equations, Eqs.~(\ref{visceqs}). In both the weak and the strong coupling cases the late-time evolution of $\Delta$ is given by the 
Navier-Stokes solution with $\Pi_{\rm NS}=4\eta/(3\tau)$. Also, note that if the pre-equilibrium evolution results in an anisotropy 
that is different from $\Delta_{\rm NS}(\tau_{\rm hydro})$, then the system relaxes to the Navier-Stokes solution within a time of 
the order of $\tau_\pi$.

As shown in Fig.~\ref{tempevol} the initial value of $\Delta$  depends on the assumed matching time.  As $\tau_{\rm hydro}$ increases, 
$\Delta_{\rm hydro}$ and $\Pi_{\rm hydro}$ increase. If the assumed value of $\tau_{\rm hydro}$ is too large then one sees an 
unreasonably fast relaxation to the Navier-Stokes solution.  This is true in the collisionally broadened scenario depicted in 
Fig.~\ref{tempevol} and also in the free-streaming scenario, $\delta=2$, which we do not explicitly plot. In the free-streaming 
scenario the longitudinal momentum-space anisotropies generated during the pre-equilibrium period are even larger.

Another issue that arises is that if the initial shear generated by the pre-equilibrium evolution becomes too large, it can become 
comparable to the equilibrium pressure ${\cal P}$.  If this is the case, then it is suspect to apply viscous hydrodynamical evolution. 
However, this is not the only possible way to generate unreasonably large shear. Once the hydrodynamical evolution begins it is 
possible to generate large shear during the integration of the hydrodynamical differential equations. This effect is larger in the 
weakly-coupling case, as the values of $\bar\eta$ and $\tau_{\pi}$ are approximately 10 and 30 times larger than in the 
strong-coupling case, respectively. This is why in Fig.~\ref{tempevol} no large values of $\Delta$ are generated in the case of 
strong coupling, whereas the weak-coupling case there are.  One other possibility that arises is that the initial value of the shear 
computed from $\xi$ will result in the initial condition being ``critical'', meaning that, when the differential equations are 
integrated, unphysical behaviors such as negative longitudinal pressures are generated \cite{Martinez:2009mf,Rajagopal:2009yw}. In our 
results, we check to see the generated initial conditions are critical and indicate wheter this happens in the corresponding results 
tables.

The evolution shown in Fig.~\ref{tempevol} is typical of the time evolution of $\Delta$ in our model.  Of course, one can vary the 
assumed value of $\xi$ at the formation time and also consider the free-streaming case.  For the sake of brevity we do not present 
plots showing these possibilities, as the analytic formulas required, Eqs.~(\ref{generalxi}), (\ref{energyaniso}), 
(\ref{xideltamatch}), and (\ref{pideltamatch}), are simple enough for readers to implement on their own. These four equations can be 
used to generate the time evolution of the plasma anisotropy for use in phenomenological applications.
In Sec.~\ref{entropyprod} we demonstrate how to use the resulting model and calculate entropy generation using it.


\section{Entropy Production} 
\label{entropyprod}

In transport theory, the entropy current is defined as \cite{GLW}
\beq
{\cal S}^\mu(x)=-\int \frac{d^3 p}{(2\pi)^3}\,\frac{p^\mu}{p^0}\,f(x,{\bf p})\bigl\{\log\bigl[f(x,{\bf p})\bigr]-1\bigr\}.
\eeq
Contracting the entropy current $S^\mu$ with the velocity fluid $u^\mu$, we obtain the entropy density 
${\cal S} \equiv u_\mu {\cal S}^\mu$.  Nonequilibrium corrections are usually computed by expanding the distribution function around 
equilibrium~\cite{GLW}. For the anisotropic distribution function, Eq.~(\ref{eq:distansatz}), the entropy density can be calculated 
analytically in the local fluid rest frame using a change of variables, giving
\bqa
\label{entropygeneral}
{\cal S} (p_{\rm hard}\,,\,\xi) &=& \frac{{\cal S}_{\rm iso} (p_{\rm hard})}{\sqrt{1+\xi}} \, ,
\eqa
which, unlike typical expressions for the nonequilibirum entropy, is accurate to all orders to $\xi$.
We note, importantly, that our ansatz, Eq.~(\ref{eq:distansatz}) does not fall into the class of distribution functions
describable using the 14 Grad's ansatz, because when expanded around equilibrium, Eq.~(\ref{eq:distansatz}) has
momentum-dependent coefficients.  Therefore, the entropy production from our anisotropic distribution will 
differ from the 14 Grad's method and IS ansatz (See Appendixes~\ref{Ap:B} and \ref{Ap:C} for a detailed comparison).
 
In both the pre-equilibrium and the viscous hydrodynamical periods we use (\ref{entropygeneral}) to calculate the percentage entropy 
generation $\Delta S/S_0$. We define
\bqa
\label{entropypercentage}
 \frac{\Delta S}{S_{0}} = \frac{\tau_{\rm fo}\,{\cal S}(\tau_{\rm fo})-\tau_0\,{\cal S}(\tau_0)}{\tau_0\,{\cal S}(\tau_0)}\equiv\frac{S_{\rm f}-S_{\rm 0}}{S_{0}} 
\eqa
where $S_{f}$ and $S_{0}$ are the entropy per unit rapidity evaluated at $\tau=\tau_{\rm fo}$ and $\tau=\tau_0$, respectively. 

Note that the two models for pre-equilibrium evolution, free-streaming and collisionally-broadened expansions, 
generate no entropy
during the pre-equilbrium period.  In the case of free-streaming it is obvious that there can be no entropy generation.  In the case of collisionally broadened expansion there is an implicit assumption that there are no inelastic processes.  Therefore,
in both cases there is no entropy generation.  This can be checked analytically by using Eqs.~(\ref{fslimit}) and (\ref{cblimit})
and computing either the entropy density or the number density, in which case one finds that both drop like $\tau^{-1}$ 
\cite{Martinez:2008di,Baier:2000sb}.\footnote{This result is found only if one uses the exact expression given by
Eq.~(\ref{entropygeneral}). If the IS or 14 Grad's expression for the nonequilibrium entropy is used, one will find that, even 
assuming a free-streaming plasma, entropy is generated during the pre-equilibrium evolution. This is obviously incorrect so we use 
Eq.~(\ref{entropygeneral}) in all cases.}
Of course, these models are an idealization and one expects inelastic processes to
contribute to entropy production during the pre-equilbrium period in a more realistic model; however, this is beyond
the scope of the current work.

The entropy produced during the expansion can be used to constrain nonequilibrium models of the QGP \cite{Dumitru:2007qr}. The 
produced entropy depends on the values of the transport coefficients and is sensitive to the assumed value of $\eta /{\cal S}$. Based 
on the fact that our pre-equilibrium models do not generate entropy, one naively expects that if the viscous hydrodynamical period 
starts later, then less entropy is produced. However, this is only true if during the viscous period we have control over the gradient 
expansion, that is, $|\Pi/{\cal P}|\ll 1$. Additionally, for fixed $p_{\rm hard}$, the factor of $\sqrt{1+\xi}$ in Eq.~(\ref{entropygeneral}) 
causes the entropy to decrease monotonically as $\xi \rightarrow \infty$. The competing effects of $p_{\rm hard}$ and $\xi$ can 
cause the naive expectation described previously to be violated, as we discuss later.

In all of the subsequent calculations shown below the parton formation time $\tau_0$ is chosen to be 0.2 fm/c and the initial 
temperature, $T_0$, at that time is taken as $0.35$ GeV.  We use a freeze-out temperature of $T_{\rm fo} = 0.16$ GeV.  
The values of the transport coefficients are summarized in Eqs.~(\ref{weaklimitvalues}) and (\ref{stronglimitvalues}). In what follows, 
we report the results for entropy production when the initial conditions are fixed at the formation time.

\subsection{Free streaming model}
\label{resultsfreestr}

\begin{figure*}[t]
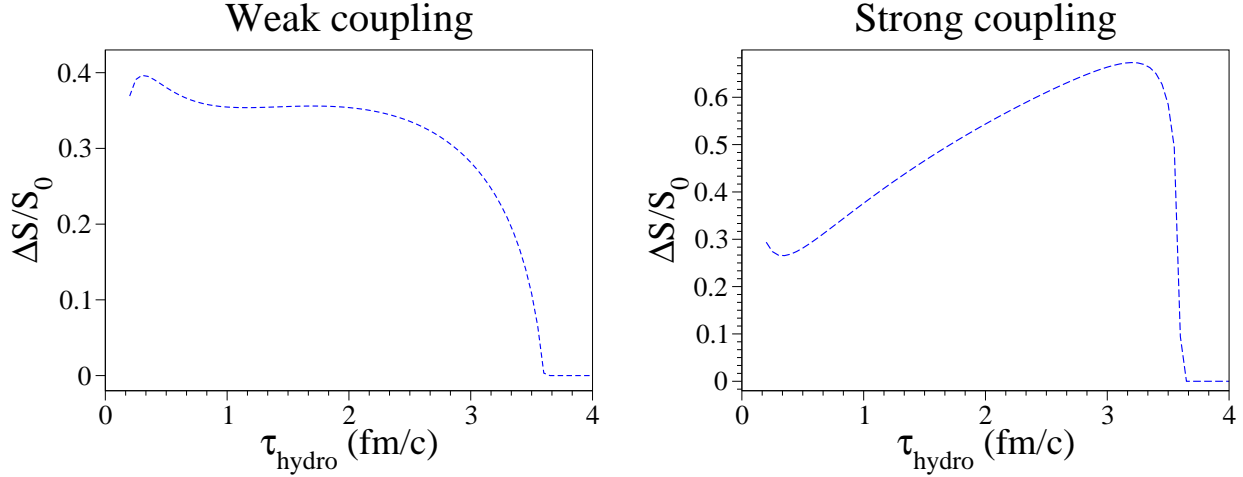

\includegraphics[scale=0.3]{3a.eps}\hspace{5mm}
\includegraphics[scale=0.3]{3b.eps}
\caption{Entropy percentage as a function of $\tau_{\rm hydro}$ for a free streaming pre-equilibrium scenario in the weak 
(left) and strong (right) coupling regimes. We use $\tau_0$ = 0.2 fm/c and the initial temperature at parton formation time $T_0$ = 0.35 GeV.}
\label{entrperc-fs}
\end{figure*}

Fig.~\ref{entrperc-fs} shows the entropy percentage as a function of $\tau_{\rm hydro}$ in the strong (left panel) and weak 
(right panel) coupling limits when free streaming is used as the pre-equilibrium scenario. In both case, after a certain time, which 
we call $\tau_c$, there is no entropy generation in our model. This is because for $\tau_{\rm hydro} > \tau_c$ the system freezes 
out while still in the pre-equilibrium period of evolution. For the free streaming model this time is approximately $\tau_c = 3.6$ 
fm/c.  

Fig.~\ref{entrperc-fs} shows that the entropy production depends on the values of the transport coefficients.
In the strong coupling case, we see that $\Delta S/S_0$ increases between $0.5\lesssim \tau_{\rm hydro}\lesssim 3$ fm/c goes to zero 
after $\tau_c$.  The increase in entropy production owes to the rapidly increasing value of $\Pi_{\rm hydro}$ in the free-streaming 
case. In the weak coupling case, there is a similar behavior but the effect is less pronounced. Again the increase in entropy 
production can be understood if one considers the values of the initial conditions $\Pi_{\rm hydro}$ and ${\cal E}_{\rm hydro}$ 
obtained from the pre-equilibrium free streaming expansion. For the free-streaming model, for example, the anisotropy parameter $\xi$ 
increases as $\tau^2$. As a consequence, one can obtain large initial values for the shear. In fact, care should be taken because the 
size of $\Pi_{\rm hydro}/{\cal P}_{\rm hydro}$ can be of $\mathcal{O}(1)$, making the use of a viscous hydrodynamical description 
after $\tau_{\rm hydro}$ suspect. Therefore, it is necessary to check the relative size of $\Pi$ and ${\cal P}$ at the matching time 
to assess the trustworthiness of the hydrodynamical evolution. The constraint on $\tau_{\rm hydro}$ is stronger if one requires 
instead a more stringent convergence criterium, $|\Pi| \leqslant {\cal P}/3$, during the hydrodynamical evolution
~\cite{Martinez:2009mf}. Here we do not apply this stronger condition but mention it to make the reader aware of this caveat.

\subsection{Collisionally broadened model}
\label{resultscollbroad}

\begin{figure*}[t]
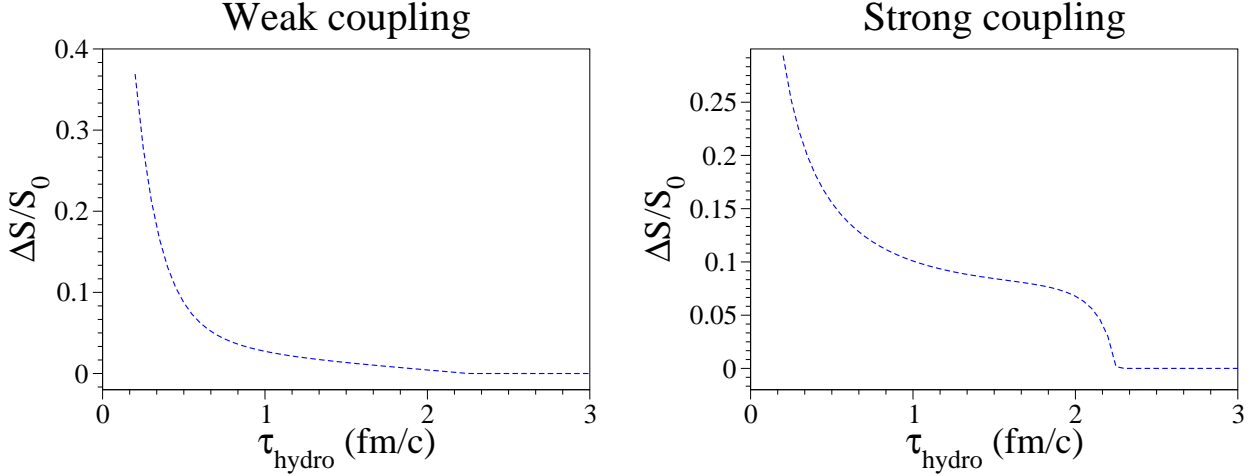

\includegraphics[scale=0.3]{4a.eps}\hspace{5mm}
\includegraphics[scale=0.3]{4b.eps}
\caption{Entropy percentage as a function of $\tau_{\rm hydro}$ for a collisionally broadening pre-equilibrium scenario in the weak 
(left) and strong (right) coupling regime. We use $\tau_0$ = 0.2 fm/c and the initial temperature at parton formation time $T_0$ = 0.35 GeV.}
\label{entrperc-cb}
\end{figure*}

In Fig.~\ref{entrperc-cb}, we show the entropy percentage $\Delta S/S_0$ in the strong (left panel) or weak (right panel) coupling 
cases as a function of $\tau_{\rm hydro}$. In the collisionally-broadened case we find $\tau_c=2.2$ fm/c, which is shorter than the 
corresponding time in the case of free streaming. This is because in the collisionally broadened (c.b.) case, the energy density 
decreases more quickly (${\cal E}_{\rm c.b}\sim \tau^{-11/9}$) than in the free streaming case 
(${\cal E}_{\rm f.s.} \sim \tau^{-1}$). 

Fig.~\ref{entrperc-cb} shows that, contrary to the case of free streaming, the entropy percentage as $\tau_{\rm hydro}$ is now a 
monotonically decreasing function of $\tau_{\rm hydro}$. This is because the initial values of the anisotropies developed 
during the pre-equilibrium period are smaller for the collisionally-broadened case than the free-streaming case~\cite{Martinez:2008di}. 
This result is more physical, as one expects that as the hydrodynamical expansion expands later (so larger $\tau_{\rm hydro}$) then 
less entropy is produced for a given value of the transport coefficients~\cite{Dumitru:2007qr}. We note that we also observe that the 
entropy percentage as a function of $\tau_{\rm hydro}$ drops more quickly in the weak coupling regime compared with the strong 
coupling case. This is a consequence of Eq.~(\ref{entropygeneral}), which shows that for fixed $p_{\rm hard}$, as $\xi$ increases, 
less entropy is produced. As pointed out in Sec.~\ref{results}, owing to the different values of the relaxation time $\tau_\pi$ in 
both coupling regimes, the anisotropy in momentum space is larger in the weak coupling case than in the strong coupling case during the viscous period~(see comparison in Fig.~\ref{tempevol}).

\subsection{Including initial anisotropies at the formation time}
\label{relaxation}

\begin{figure*}[t]
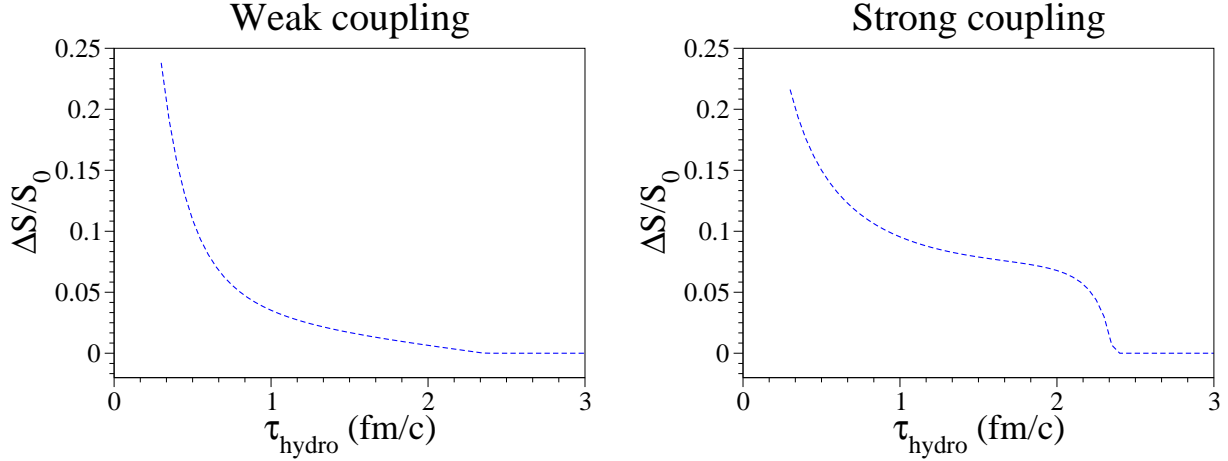

\includegraphics[scale=0.29]{5a.eps}\hspace{5mm}
\includegraphics[scale=0.29]{5b.eps}
\caption{Entropy percentage as a function of $\tau_{\rm hydro}$ for a collisionally broadening pre-equilibrium scenario and 
$\xi_0$=-0.1, in the strong (right) and weak (left) coupling regime. We use $\tau_0$ = 0.2 fm/c and the initial 
temperature at parton formation time $T_0$ = 0.35 GeV.}
\label{entrperc-cb-xi-0.1}
\end{figure*}

In the previous subsections we reported results for the case where there was no momentum-space anisotropy at parton formation time, 
that is, $\xi_0=0$.  Here we relax this assumption.
In Fig.~\ref{entrperc-cb-xi-0.1}, we show the result for entropy production for a prolate initial distribution with
$\xi_0=-0.1$ in the collisionally-broadened scenario. In both coupling cases, $\Delta S /S_0$ decreases as $\tau_{\rm hydro}$ 
increases. Because the initial value of the anisotropy is close to zero, generally speaking, the behavior of the entropy percentage 
is similar to the case where there is an isotropic initial state ($\xi_{0}=0$).

For larger values of $\xi_0$, the situation becomes more complicated because we do not have control of the size of $\Pi/{\cal P}$ and 
the system can become ``critical''.  Therefore, for extreme initial anisotropies, it is not possible to determine $\tau_{\rm hydro}$ 
based on entropy constraints. In Table~\ref{tauconstraint}, we summarize the bounds on $\tau_{\rm hydro}$ obtained by varying $\xi_0$ 
in the strong and weak coupling regimes when one fixes the initial conditions using collisionally broadening expansion.  
``Non determined'' indicates cases in which the initial anisotropies were so extreme as to cause the system to begin
generating negative longitudinal pressure.  In those cases the viscous hydrodynamics is unreliable.
The striking conclusion from Table~\ref{tauconstraint} is that, in all cases, the lower bound on $\tau_{\rm hydro}$ owing solely to 
entropy considerations is larger in the weak coupling case than in the strong coupling case. This is caused by the competing effect 
between increasing anisotropy and dropping temperature in Eq.~(\ref{entropygeneral}) and can be seen in the fact that the entropy 
production decreases more rapidly in the weak coupling panels in Figs.~\ref{entrperc-cb} and \ref{entrperc-cb-xi-0.1}.  We point out 
that these estimates are lower bounds on the minimum $\tau_{\rm hydro}$, as our models have no entropy generation during the 
pre-equilibrium period.  In a more realistic scenario there would also be entropy generation during the pre-equilibrium period, 
which would add to all curves presented in this section.

In closing we emphasize that the bounds in Table~\ref{tauconstraint} do not factor in the constraint that the shear should be small 
compared to the isotropic pressure.  As shown in Ref.~\cite{Martinez:2008di} requiring $\Pi/{\cal P} < 1/3$ as a convergence criterion 
for viscous hydrodynamics and assuming an initially isotropic plasma (i.e., $\Pi_{\rm hydro}=0$) result in a constraint 
$\tau_{\rm hydro} > 5.9 \, T_0^{-1}$ in the case of a weakly coupled plasma and $\tau_{\rm hydro} > 0.85 \, T_0^{-1}$ in the case of 
a strongly coupled plasma.  Assuming an initial temperature of 350 MeV this gives $\tau_{\rm hydro} > 3.3$ fm/c in the case of a 
weakly coupled plasma and $\tau_{\rm hydro} > 0.6$ fm/c in the case of a strongly coupled plasma.  Therefore, assuming an initially 
isotropic plasma, the convergence constraint can be than the constraint implied by entropy production. In general one must compare 
both constraints to determine which results in a stronger condition.

\begin{table}[t]
\begin{tabular}{ | c || c | c | c |}\hline
\multicolumn{3}{| c |}{$\Delta S/S_0 \leq$ 10\%}\\
\hline\hline
{\bf $\; \xi_0 \;$} & 
{\bf $\;$ Weak coupling $\;$ } & 
{\bf $\;$ Strong coupling$ \;$} \\ \hline
-0.5 & $\tau_{\rm hydro} \geq$ 0.65 fm/c  & $\tau_{\rm hydro} \geq$ 0.75 fm/c \\ \hline
0 & $\tau_{\rm hydro} \geq$ 0.45 fm/c  & $\tau_{\rm hydro} \geq$ 0.9 fm/c \\ \hline
10 &  Non determined &  $\tau_{\rm hydro} \geq$ 0.75 fm/c \\ \hline
\end{tabular}
\begin{tabular}{ | c || c | c | c |}\hline
\multicolumn{3}{| c |}{$\Delta S/S_0 \leq$ 20\%}\\
\hline\hline
{\bf $\; \xi_0 \;$} & 
{\bf $\;$ Weak coupling $\;$ } & 
{\bf $\;$ Strong coupling$ \;$} \\ \hline
-0.5 &  Non determined  &  Non determined \\ \hline
0 & $\tau_{\rm hydro} \geq$ 0.3 fm/c  & $\tau_{\rm hydro} \geq$ 0.35 fm/c \\ \hline
10 & Non determined &  $\tau_{\rm hydro} \geq$ 0.65 fm/c \\ \hline
\end{tabular}
\caption{Bounds on $\tau_{\rm hydro}$ imposed by requiring either a 10\% (left) or 20\% (right) bound on percentage 
entropy when considering different values of $\xi_0$ and transport coefficients. We fix the initial conditions through  0+1 
collisionally broadening expansion. $\tau_0$ = 0.2 fm/c and $T$= 350 MeV.}
\label{tauconstraint}
\end{table}


\section{Conclusions and Outlook}
\label{conclusions}  

In this paper we have presented a model that allows us to match 0+1 pre-equilibrium dynamics and 0+1 second-order conformal viscous 
hydrodynamics at a specified proper-time $\tau_{\rm hydro}$. The pre-equilibrium evolution is modeled by either free-streaming or 
collisionally-broadening expansion. We have derived a relation between the microscopic anisotropy parameter $\xi$ and the pressure 
anisotropy of the fluid $\Delta$.  This relation allowed us to determine the initial conditions for the shear $\Pi$ and energy density 
${\cal E}$ at the assumed matching time.  The initial values of ${\cal E}$ and $\Pi$ depend on the kind of pre-equilibrium model 
considered and, also, on the interval of time over which pre-equilibrium dynamics is assumed to take place. The resulting models can 
be used to assess the impact of pre-equilibrium dynamics on a variety of observables such as photon and dilepton production, 
heavy-quark transport, jet-medium induced electromagnetic radiation, etc.

As a particular application here we have studied entropy generation as a function of $\tau_{\rm hydro}$. We have derived an exact 
expression for the nonequilibrium entropy and then used this to determine the percentage entropy generation.  We have shown that 
owing to the reduction in entropy by a factor of $\sqrt{1+\xi}$ compared to the isotropic case, it is possible to have more entropy 
generation in the strongly coupled case.  We have summarized our results for entropy generation in 
Table~\ref{tauconstraint} by presenting bounds on 
$\tau_{\rm hydro}$ that result from requiring the percentage entropy generation to be less than 10\% or less than 20\% .

It is possible to extend these studies to higher dimensions and match all components of the energy-momentum tensor and fluid 
four-velocity. To do this it is necessary to specify information about the transverse expansion during the pre-equilibrium period and 
how this impacts the anisotropy at early times. One way to approach this problem is to use three dimensional (3D) parton cascade 
models~\cite{Xu:2004mz,Xu:2007aa,Huovinen:2008te} or 3D Boltzmann-Vlasov-Yang-Mills simulations \cite{Dumitru:2006pz}. Additionally, 
in our approach there is no entropy generation during the pre-equilibrium phase. This is not necessarily true, as inelastic 
collisions such as $2\to 3$ are necessary for chemical equilibration~\cite{Baier:2000sb,Xu:2004mz,Xu:2007aa} and therefore, their 
inclusion will produce entropy during the nonequilibrium phase of the QGP. Short of this, one can investigate simple analytic models 
such as 3D free-streaming or 3D collisionally-broadened expansion and develop analytic models that can be used to determine the 
necessary initial conditions self-consistently.  Work along these lines is currently in progress.


\section*{Acknowledgements}

We thank A. Dumitru, G. Denicol, A. El, M. Gyulassy, P. Huovinen, J. Noronha, P. Romatschke, D. Rischke and B. Schenke for useful 
discussions and suggestions. M. Martinez thanks the Physics Department at Gettysburg College for the kind hospitality where part of 
this work was done. M. Martinez is grateful to R. Loganayagam for the explanation of his work
~\cite{Loganayagam:2008is, Bhattacharyya:2008xc}. M. Martinez was supported by the Helmholtz Research School and Otto Stern School of 
the Goethe-Universit\"at Frankfurt am Main. M. Strickland was supported partly by the Helmholtz International Center for FAIR 
Landesoffensive zur Entwicklung Wissenschaftlich-\"Okonomischer Exzellenz program.

\appendix
\section{Landau matching conditions of the anisotropic distribution function during the viscous period}
\label{Ap:A}
To determine the isotropic equilibrium energy density ${\cal E}_{\rm eq}(T)$ from a nonequilibrium single-particle distribution 
function, it is necessary to implement the Landau matching conditions
\begin{subequations}
\label{Landauconstrains}
\begin{align}
\label{energyconstrain}
{\cal E}(T)&=u_\mu T_{(0)}^{\mu\nu}u_\nu \, , \\
\label{deltaNconstrain}
u_\mu\delta N^\mu &=0\, ,\\
\label{deltaTconstrain}
u_\mu\delta T^{\mu\nu}u_\nu&=0 \, , 
\end{align}
\end{subequations}
where $T_{(0)}^{\mu\nu}$ is the energy-momentum tensor computed with the equilibrium distribution function $f_{\rm eq}(x,p)$, and 
$\delta T^{\mu\nu}$ involves nonequilibrium corrections to the energy-momentum tensor. In the 14 Grad's method this constraint is 
immediately satisfied by construction. In the case of the anisotropic Boltzmann distribution, the first constraint, 
Eq.~(\ref{energyconstrain}), requires
\beq
\int \frac{d^3{\bf p}}{(2\pi)^3 p^0}(u\cdot p)^2 \exp \bigl[-\sqrt{{\bf p^2}+\xi\, p_z^2}\,/\,p_{\rm hard}\bigr] = \int \frac{d^3{\bf p}}{(2\pi)^3 p^0}(u\cdot p)^2 \exp \bigl[- p /T\bigr]\,. 
\eeq
Performing the integrals on both sides, we find that

\beq
\label{phardconstrain}
p_{\rm hard}= \bigl({\cal R} (\xi)\bigr)^{-1/4}\,T\, .
\eeq
Now we expand the anisotropic distribution function to second order in $\xi$, and making use of the last expression, we have
\footnote{For practical purposes we expand until second order in $\xi$, as we are considering viscous hydrodynamics until second 
order in gradient expansion.}
\bqa
f_{\rm aniso}({\bf p},\xi,T)&=&\exp \bigl[-\sqrt{{\bf p^2}+\xi\, p_z^2}\,/\,p_{\rm hard}\bigr]\\ \nonumber
&=& \exp \biggl[-\frac{p}{T}\bigl({\cal R} (\xi)\bigr)^{1/4}\,\sqrt{1+\xi\,\cos \theta}\biggr]\\ \nonumber
&\approx& e^{-p/T}\bigl(1+\xi f_{(1)}+\xi^2 f_{(2)}\bigr),
\eqa
where we use explicitly $p_z= p \cos\theta$. The functions $f_{(1)}$ and $f_{(2)}$ are given by
\begin{subequations}
\label{xicorrections}
\begin{align}
\label{xicorrection}
f_{(1)}&=\frac{p}{6\,T}(1-3\cos^2\theta)
\\
\label{xi2correction}
f_{(2)}&=\frac{p}{360\,T^2}\bigl[5p-39T+30(T-p)\cos^2\theta+45(T+p)\cos^4\theta\bigr]\,.
\end{align}
\end{subequations}
Replacing the expansion of the anisotropic distribution until ${\cal O}(\xi)$ in the Landau condition, Eq.~(\ref{deltaNconstrain})
\bqa
u_\mu\delta N^\mu&=&\xi \,\int \frac{d^3p}{(2\pi)^3}\,e^{-p/T}\,f_{(1)}\\ \nonumber
&=&\frac{1}{(2\pi)^2}\frac{1}{6T}\,\int_0^\infty\int_0^\pi dp\,d(\cos\theta)\,p^3\,e^{-p/T}\,(1-3\cos^2\theta)\\ \nonumber
&=&0\,.
\eqa
To ${\cal O}(\xi)$, the Landau condition, Eq.~(\ref{deltaNconstrain}), is satisfied. Note that this condition is not expected to hold 
at  all orders because the particle number density is proportional to the entropy density. The only way to hold both the energy 
density and the particle number density is by introducing a chemical potential.

Expanding the anisotropic distribution to ${\cal O}(\xi^2)$ in the Landau condition~(\ref{deltaTconstrain}), we have
\beq
u_\mu\delta T^{\mu\nu}u_\nu= \int \frac{d^3p}{(2\pi)^3}\,p\,e^{-p/T}\,(\xi f_{(1)}+\xi^2 f_{(2)})\,,
\eeq
where $f_{(1)}$ and $f_{(2)}$ are given by Eqs.~(\ref{xicorrections}). Explicitly, we have
\bqa
\label{O(xi)}
\xi\int \frac{d^3p}{(2\pi)^3}\,p\,e^{-p/T} f_{(1)}&=& \frac{\xi}{(2\pi)^36T}\int d^3p\,p^2\,e^{-p/T}\bigl(1-3\cos^2\theta\bigr)\\ 
&=& \frac{\xi}{(2\pi)^26T}\int dp\,p^4\,e^{-p/T}\int d(\cos\theta)\bigl(1-3\cos^2\theta\bigr)\\ \nonumber
&=&0 \,,\nonumber
\eqa

\bqa
\label{O(xi2)}
\xi^2\int \frac{d^3p}{(2\pi)^3}\,p\,e^{-p/T}f_{(2)}&=& \frac{\xi^2}{(2\pi)^3\,360\,T^2}\int d^3p\,p^2\,e^{-p/T}\bigl(5p-39T
\\ \nonumber
&+&30(T-p)\cos^2\theta+ 45(p+T)\cos^4\theta\bigr)\\ \nonumber
&=&\frac{\xi^2}{(2\pi)^2\,360\,T^2}\biggl(8\int_0^\infty dp\, p^5\,e^{-p/T}-40\,T\int_0^\infty dp\, p^4\,e^{-p/T}\biggr)\\ \nonumber
&=&0 \,.\nonumber
\eqa
Therefore, up to second order in $\xi$, the anisotropic distribution function satisfies the Landau condition, Eq.~(\ref{deltaTconstrain}). 


\section{Entropy from the 14th Grad's Method}
\label{Ap:B}

We can evaluate the entropy from the kinetic theory definition using the 14th Grad's approximation for the nonequilibrium 
distribution function
\beq
\label{fexpansion}
f(x,p)= f_{\rm eq}(1+\delta f),
\eeq
The dependence of $\delta f$ is assumed to be a function of the hydrodynamic degrees of freedom 
${\cal E}, {\cal P}, u^\mu, g^{\mu\nu}$, and $\pi^{\mu\nu}$ expanded in a Taylor series
\beq
\delta f(x^\mu, p^\mu)= \epsilon_0 + \epsilon_\mu p^\mu + \epsilon_{\mu\nu}p^\mu p^\nu + {\cal O} (p^3).
\eeq
By demanding that $\delta f$ vanishes in equilibrium, one finds that in the Landau frame and assuming massless Boltzmann particles, $\delta f$ is given by~\cite{Israel:1976tn}
\beq
\label{14gradansatz}
\delta f(x^\mu,p^\mu)=\frac{1}{2\, T^2\,({\cal E}(T)+{\cal P}(T))}\,\pi_{\mu\nu} p^\mu p^\nu + {\cal O}(p^3)
\eeq
Expanding the expression of the entropy
\bqa
{\cal S}&=&- \int \frac{d^3p}{(2\pi^3)}\,f_{\rm eq}(1+\delta f)\bigl[\log (f_{\rm eq}(1+\delta f))-1\bigr]\\ \nonumber
&\approx& {\cal S}_{(0)}+ {\cal S}_{(1)}+{\cal S}_{(2)}.
\eqa
where
\begin{subequations}
\begin{align}
{\cal S}_{(0)}&=-\int\frac{d^3p}{(2\pi^3)}f_{\rm eq}\bigl[\log (f_{\rm eq})-1\bigr],\\
{\cal S}_{(1)}&=-\int\frac{d^3p}{(2\pi^3)}f_{\rm eq}\,\delta f\,\log [f_{\rm eq}],\\
{\cal S}_{(2)}&=-\frac{1}{2}\int\frac{d^3p}{(2\pi^3)}f_{\rm eq}\,(\delta f)^2\,.
\end{align}
\end{subequations}

After replacing the 14 Grad's ansatz in the last expressions, these integrals can be calculated analytically if one rewrites them as moments of the equilibrium distribution function $f_{\rm eq}(x^\mu,p^\mu)= e^{-p/\,T}$. After a lengthy calculation, we find:
 \begin{subequations}
\begin{align}
{\cal S}_{(0)}&= \frac{1}{T}\biggl(\frac{3\,T^4}{\pi^2}+\frac{T^4}{\pi^2}\biggr)\equiv\frac{1}{T}\bigl({\cal E}(T)+{\cal P}(T)\bigr) \\
{\cal S}_{(1)}&= 0\\,
{\cal S}_{(2)}&=-\frac{3}{8}\,\frac{\pi_{\mu\nu}\pi^{\mu\nu}}{T\,{\cal P}(T)}
\end{align}
\end{subequations}
Using the ideal equation of state, the nonequilibrium entropy is 
\beq
\label{nonequilentr}
{\cal S}_{\rm noneq}= \frac{4}{3\,T}{\cal E}(T)-\frac{3}{8}\frac{\pi_{\mu\nu}\pi^{\mu\nu}}{T\,{\cal P}(T)} \, .
\eeq
Comparing this expression with the IS ansatz for the nonequilibrium entropy, Eq.~(\ref{viscentropy}), we find a well-known result for a Boltzmann gas
\beq
\label{beta2}
\beta_2= \frac{3}{4\,{\cal P}(T)}\,\,.
\eeq
For 0+1-dimensional case, where $\pi_{\mu\nu}= {\rm diag}\,(0,\Pi/2,\Pi/2,-\Pi)$, the nonequilibrium entropy, 
Eq.~(\ref{nonequilentr}), is
\beq
\label{0+1nonequilentr}
{\cal S}_{\rm noneq}=\frac{4}{3\,T}{\cal E}(T)-\frac{3}{4}\frac{\beta_2}{T}\Pi^2\,.
\eeq
%

\section{Entropy from the anisotropic distribution ansatz}
\label{Ap:C}

Using the kinetic theory framework, we can calculate the entropy from the anisotropic distribution ansatz, Eq. ~(\ref{eq:distansatz}), 
expanding in a Taylor series in terms of the anisotropy parameter $\xi$ (Eqs.~\ref{xicorrections}). Since the Landau matching 
conditions are satisfied by the anisotropic distribution function, one can write the shear tensor to first order in $\xi$ as
\bqa
\label{pikinetic}
\Pi^{\mu\nu}&=& \int \frac{d^3p}{(2\pi)^3p^0}p^\mu p^\nu f_{eq}\delta f \\ \nonumber
&=&\frac{\xi}{6\,T}\int \frac{d^3p}{(2\pi)^3}\,p^\mu p^\nu\,e^{-p/T}\,(1-3\cos^2\theta)\, ,
\eqa
where we use explicitly the first-order correction to the anisotropic distribution function, Eq.~(\ref{xicorrection}). We are 
interested in the 0+1-dimensional case, where there is just one independent component of the shear tensor $\Pi^{zz}=-\Pi$. Calculating the $zz$ component from the last expression, we have
\bqa
\label{Pizz}
\pi^{zz}\equiv -\Pi&=&\frac{\xi}{6\,T}\int \frac{d^3p}{(2\pi)^3}\,p_z^2\,e^{-p/T}\,(1-3\cos^2\theta)\, ,\\ \nonumber
&=& \frac{\xi}{(2\pi)^2\,6\,T}\int_0^\pi d(\cos\theta)\cos^2\theta\,(1-3\cos^2\theta)\,\int_0^\infty dp\,p^4\,e^{-p/T}\, ,\\ \nonumber
&=& -\frac{8}{15}\frac{T^4}{\pi^2}\xi\, ,\\ \nonumber
&=&-\frac{8}{45}\,{\cal E}(T)\,\xi\,.
\eqa
This expression coincides with Eq.~(\ref{smallxiviscous}).

$\,$\\

Expanding the entropy  to second order in $\xi$:
\bqa
\label{0+1viscentropyapprox}
{\cal S}_{noneq}&=&-\int \frac{d^3p}{(2\pi)^3}f_{\rm eq}\bigl[1+\xi f_{(1)}+\xi^2 f_{(2)}\bigr]\bigl[\log\bigl[f_{\rm eq}\bigl(1+\xi f_{(1)}+\xi^2 f_{(2)}\bigr)\bigr]-1\bigr]\, ,\\ \nonumber
&\approx&{\cal S}_{(0)}+ {\cal S}_{(1)}+{\cal S}_{(2)}.
\eqa
where
\bqa
{\cal S}_{(0)}&=& -\int \frac{d^3p}{(2\pi)^3}f_{\rm eq}\bigl[\log f_{\rm eq}-1\bigr]\, , \\ 
{\cal S}_{(1)}&=&-\xi\int \frac{d^3p}{(2\pi)^3}f_{(1)} f_{\rm eq}\log f_{\rm eq}\, ,\\ 
{\cal S}_{(2)}&=&-\xi^2\int \frac{d^3p}{(2\pi)^3}f_{\rm eq}\biggl(\frac{(f_{(1)})^2}{2}+f_{(2)}\log f_{\rm eq}\biggr)\,.
\eqa
After replacing $f_{(1)}$ and $f_{(2)}$ (Eqs.~(\ref{xicorrections})) in the last expressions, and with $f_{\rm eq}= e^{-p/T}$, we have
\bqa
{\cal S}_{(0)}&=& \frac{1}{T}\biggl(\frac{3\,T^4}{\pi^2}+\frac{T^4}{\pi^2}\biggr)\equiv\frac{1}{T}\bigl({\cal E}(T)+{\cal P}(T)\bigr)\\ 
{\cal S}_{(1)}&=& 0 \\
{\cal S}_{(2)}&=& -\frac{2}{15\,T}\frac{T^4}{\pi^2}\xi^2\\ \nonumber
&=&-\frac{2}{45}\frac{{\cal E}(T)}{T}\xi^2\,.
\eqa
Using the ideal equation of state, the nonequilibrium entropy can be written as
\bqa
{\cal S}_{\rm noneq}&=&\frac{4}{3}\frac{{\cal E}(T)}{T}-\frac{2}{45}\frac{{\cal E}(T)}{T}\xi^2\,  \\\nonumber
&=&\frac{4}{3}\frac{{\cal E}(T)}{T}\biggl(1-\frac{\xi^2}{30}\biggr)\\\nonumber
&=&\frac{4}{3}\frac{{\cal E}(T)}{T}\biggl(1-\frac{135}{128}\,\biggl(\frac{\Pi}{{\cal E}(T)}\biggr)^2\biggr)\,.
\eqa
If one compares this result with the IS ansatz for the nonequilibrium entropy in the 0+1-dimensional case, Eq.~(\ref{0+1nonequilentr}), the term $\beta_2$ for the anisotropic distribution ansatz can be fixed as
\beq
\label{beta2aniso}
\beta_2=\frac{5}{8}\frac{1}{{\cal P}(T)}
\eeq
Note that the difference in the values between Eqs.~(\ref{beta2}) and Eq.~(\ref{beta2aniso}) comes from the fact that the anisotropic 
distribution is incompatible with the 14 Grad's ansatz.  This is because at small $\xi$ the linear term 
coming from the ansatz, Eq.~(\ref{eq:distansatz}), cannot be expressed in the form $a_{42} \, p_\mu \Pi^{\mu\nu}p_\nu $,
 with $a_{42}$ being a momentum-independent quantity.  In the case of Eq.~(\ref{eq:distansatz}) the corresponding coefficient of 
$p_\mu \Pi^{\mu\nu}p_\nu$ is momentum dependent, and hence Eq.~(\ref{eq:distansatz})  does not fall into the class of distribution functions describable using the 14 Grad's ansatz.


\bibliography{initialcond}

\end{document}